%% file: main.tex
\documentclass[letterpaper,twocolumn,10pt]{article}
\usepackage{usenix-2020-09}
\usepackage[T1]{fontenc}
\usepackage{algorithm}
\usepackage[noend]{algpseudocode}
\usepackage{amsmath,amsthm,mathtools}
\usepackage{amssymb}
\usepackage{cleveref}
\usepackage{thmtools}
\usepackage{listings}
\usepackage{booktabs}
\usepackage{nicematrix}
\usepackage{xspace}
\usepackage[detect-all]{siunitx}
\usepackage{paralist}
\usepackage{multirow}
\usepackage{inconsolata}
\usepackage{tikz}
\usepackage{pifont}
\usepackage{graphicx}
\usetikzlibrary{positioning,backgrounds}

\usepackage[available]{usenixbadges}

\newcommand{\ra}[1]{\renewcommand{\arraystretch}{#1}}

\usepackage[labelfont={bf}]{caption}

\usepackage{enumitem}
\newlist{EV}{enumerate}{1}
\setlist[EV]{label=($V_{\arabic*}$),ref=$V_{\arabic*}$}
\newlist{EQ}{enumerate}{1}
\setlist[EQ]{label=($Q_{\arabic*}$),ref=$Q_{\arabic*}$}

\usepackage{float}
\newfloat{algorithm}{t}{lop}

\newfloat{listing}{tbp}{lol}
\floatname{listing}{Listing}

\newcommand{\name}{Ote\xspace}
\newcommand{\diaspora}{diaspora\xspace}

\newcommand{\prune}{\reflectbox{\ding{36}}}

\crefname{section}{\S}{\S\S}
\crefformat{section}{\S~#2#1#3}
\Crefformat{section}{\S~#2#1#3}

\DeclareMathOperator{\arity}{arity}
\DeclareMathOperator{\dom}{dom}

\newtheorem{theorem}{Theorem}[section]

\newtheorem{fact}[theorem]{Fact}

\theoremstyle{definition}

\newtheorem{example}[theorem]{Example}
\theoremstyle{plain}

\theoremstyle{remark}

\theoremstyle{plain}

\AtBeginEnvironment{example}{%
  \pushQED{\qed}%
}
\AtEndEnvironment{example}{\popQED\endexample}

\setlist{noitemsep,parsep=0pt,topsep=1pt,partopsep=1pt}

\lstnewenvironment{compactsql}{\lstset{language=sql,aboveskip=0pt,belowskip=0pt,xleftmargin=0em}}{}

\def\tlathin{\mskip.5\thinmuskip}
\newcommand{\tupL}{\ensuremath{\langle\tlathin}}
\newcommand{\tupR}{\ensuremath{\tlathin\rangle}}

\begin{document}

\date{}

\title{\Large \bf Extracting Database Access-Control Policies From Web Applications}

\author{
{\rm Wen Zhang\thanks{Now at Google.}}\\
UC Berkeley
\and
{\rm Dev Bali}\\
UC Berkeley
\and
{\rm Jamison Kerney}\\
UC Berkeley
\and
{\rm Aurojit Panda}\\
NYU
\and
{\rm Scott Shenker}\\
UC Berkeley and ICSI
} 

\maketitle

\begin{abstract}
\input{abstract}
\end{abstract}

\section{Introduction}
\input{intro}

\section{Motivation and Background}
\input{setting}

\section{Overview}\label{sec:overview}
\input{overview}

\section{Exploring Executions}\label{sec:concolic}
\input{exploration}

\section{Generating a Policy}\label{sec:policy}
\input{policy}

\section{Implementation and Practical Aspects}\label{sec:implementation}
\input{implementation}

\section{Evaluation}\label{sec:eval}
\input{eval}

\section{Related Work}\label{sec:related}
\input{related}

\section{Discussion and Future Work}
\input{conclusion}

\section*{Acknowledgments}
We thank the anonymous reviewers and members of the UC Berkeley NetSys Lab and Sky Computing Lab for their feedback.
This work is supported in part by NSF grant~2145471 and by gifts from Accenture, Algorithmic SuperIntelligence Labs, Amazon, AMD, Anyscale, Broadcom, cmpnd, Google, IBM, Intel, Intesa Sanpaolo, Lambda, Lightspeed, NVIDIA, Samsung SDS, and SAP.

\bibliographystyle{plain}
\bibliography{main}

\clearpage
\appendix
\input{artifact-appendix}

\end{document}

%% file: abstract.tex
To safeguard sensitive user data,
web developers typically rely on implicit access-control policies, which they implement
using access checks and query filters.
This ad hoc approach is error-prone as these scattered checks and filters are easy to misplace or misspecify,
and the lack of an explicit policy precludes external access-control enforcement.
More critically,
it is difficult for humans to discern what policy is embedded in application code (i.e., what data the application may access)---%
an issue that worsens as development teams evolve.

This paper tackles \emph{policy extraction}: the task of extracting the access-control policy embedded in an application by summarizing its data queries.
An extracted policy, once vetted for errors, can stand alone as a specification for the application's data access, and can be enforced to ensure compliance as code changes over time.
We introduce \name, a policy extractor for Ruby on Rails web applications.
\name uses concolic execution to explore execution paths through the application, generating traces of SQL queries and conditions that trigger them.
It then merges and simplifies these traces into a final policy that aligns with the observed behaviors.
We applied \name to three real-world applications and compared extracted policies to handwritten ones, revealing several errors in the latter.

%% file: intro.tex
Protecting sensitive data from unauthorized access is a critical concern for today's
web applications.
Therefore, when building web applications, developers must determine what \emph{access-control policy} the application should enact---%
for example, a university might want a policy that ensures a student's grades are visible only to the student and their instructors.

In today's applications, access-control policies are embedded in application code.
Furthermore, in most cases they are spread across several functions and in the filter predicates of multiple database queries.
This practice is error-prone: Missing or misspecified access checks have previously led to sensitive-data exposure~\cite{stock18:hotcrp_hidden_tags,kohler13:hotcrp_hide_rounds,kohler15:hotcrp_download_pc,ashford15:facebook_photo,maunder16:wordpress,green17:piazza}.
But more fundamentally, because the policy is never stated explicitly, it is difficult for anyone other than the application's developer to understand \emph{what} policy is embedded in the code.
Worse, as time passes, even the application development team is unlikely to remember the policy, and is unlikely to be able to reconstruct it from application code.
While there have been research frameworks that require explicit policy specification~\cite{DBLP:conf/popl/YangYS12,DBLP:conf/pldi/YangHASFC16,DBLP:conf/pldi/AustinYFS13,DBLP:conf/sosp/AlbabAATPRS24,DBLP:conf/osdi/AdamZZRHJCKS25}, we aim to address legacy applications rather than requiring them to be rewritten in such frameworks.

This paper tackles the task of \emph{policy extraction}: extracting a web application's implicitly embedded access-control policy by summarizing its possible data accesses.
A human then reviews the extracted policy to better understand the application's data accesses, ensuring they are within the bounds of intended data revelations.
If not, the application likely has an access-check bug to be fixed.
Once reviewed, the policy can stand alone as a specification for the application's data accesses and can optionally be enforced using an enforcer~\cite{DBLP:conf/uss/MehtaEH0D17,DBLP:conf/hotos/MarzoevASYKMKM19,DBLP:conf/osdi/ZhangSCPSS22} to ensure continued compliance.

We present an approach for extracting policies from legacy web applications.
Our approach begins by exploring execution paths through application code using concolic execution~\cite{DBLP:conf/pldi/GodefroidKS05,DBLP:conf/sigsoft/SenMA05}, producing transcripts that record the conditions under which SQL queries are issued~(\Cref{sec:concolic}).
These transcripts are then merged and simplified to derive a policy
that allows each recorded query to be issued under its conditions~(\Cref{sec:policy}).

A key challenge here is scalability: Web-backend code is often complex with many branches, making exhaustive path exploration infeasible.
But we empirically observe that the logic governing query issuance typically relies on only a small set of simple operations~(\Cref{sec:concolic:simple}).
We thus tailor concolic execution to track only those operations, reducing the path space (and implementation effort).
We further prune exploration using an LLM-based relevance judge (\Cref{sec:concolic:pruning}) that identifies and ignores branches unrelated to data access, reducing exploration time from potentially days to mere hours.

We implemented this approach in \name, a policy-extraction tool for web applications written in Ruby on Rails.
We then applied \name to three real-world applications, two of which we had previously written policies for by hand~\cite{DBLP:conf/osdi/ZhangSCPSS22}.
When we compared the extracted policies to the handwritten ones~(\Cref{sec:eval:compare}), we identified several formerly unknown errors in the latter---%
including a few overly permissive views that reveal sensitive data to unauthorized users.
This underscores the difficulty of understanding access-control logic in complex legacy applications and the utility of policy extraction in aiding this understanding.
A further review of the extracted policies uncovered a subtle bug we had inadvertently introduced into application code that silently disabled an access check.
These findings show that extracting and then inspecting policies results in a clearer understanding of what the program does---and should do---than current practice provides.

Due to its use of concolic execution and an LLM relevance judge, \name cannot guarantee that the extracted policy covers all possible queries or captures every condition under which queries are issued~(\Cref{sec:overview:scope}).
\name also needs the user to specify certain input constraints~(\Cref{sec:overview:workflow}) and to review the relevance judge's results~(\Cref{sec:concolic:pruning}),
supports only a subset of SQL~(\Cref{sec:overview:scope}),
and could use a better user interface for policy auditing.
Even so, our evaluation~(\Cref{sec:eval}) demonstrates that for real-world applications, \name extracts policies that are immediately useful,
identifying several errors in handwritten policies and application code.
These results show that \name already provides concrete, practical value and represents a meaningful step toward improving data security in existing web applications.

%% file: setting.tex
\subsection{Challenges of Policy Creation}\label{sec:setting:challenges}
We were prompted to tackle the policy-extraction problem by our earlier experience hand-crafting policies for existing web applications.
A few years ago, while investigating externally enforced access control for web applications,%
\footnote{By ``externally enforced'', we mean having an enforcer program mediate the application's database queries, preventing accesses deemed unauthorized according to an access-control policy~\cite{DBLP:conf/osdi/ZhangSCPSS22,DBLP:conf/hotos/MarzoevASYKMKM19,DBLP:conf/uss/MehtaEH0D17,DBLP:conf/vldb/LeFevreAERXD04}.}
we took open-source applications,
chose several representative URL endpoints, and tried our best to write down policies allowing the data accesses needed for their intended function~\cite[\S~8.1]{DBLP:conf/osdi/ZhangSCPSS22}.

This process was extremely tedious and time-consuming.
We reviewed documentation, inspected database schemas, experimented with the applications using sample data, and read their source code.
Using this information and common sense, we formulated policies that we thought would allow the endpoints to function correctly while protecting sensitive data.
Despite our best efforts, we often discovered errors in our drafts, including an omission that would have leaked sensitive data---one that we only later discovered by chance.

Our struggle was not unique:
Policy creation is known as a major challenge across security domains including role-based~\cite[\S~6.5]{rbac-nist}, attribute-based~\cite[p.~39]{DBLP:conf/birthday/DasMAVS18}, and relationship-based~\cite{DBLP:conf/sacmat/BuiSL19} access control, securing cloud resources~\cite{DBLP:conf/cav/Rungta22}, and sandboxing syscalls~\cite{DBLP:journals/pacmpl/Pailoor0SD20}.
While solutions have been proposed for some domains (\Cref{sec:related}), generating database policies for web applications has remained an open problem that is uniquely challenging, due to both the dynamism of web applications and the fine granularity of the policies required.

\subsection{Policy as SQL View Definitions}
Before delving into how \name extracts policies, we first describe how our access-control policies are specified.

We target web applications that store data in a relational database; when a user visits a page, the application issues SQL queries on the user's behalf and renders the page using the query results.
In this context, a classic way to specify access-control policies is to use a list of \emph{view definitions}~\cite{DBLP:conf/icde/Motro89,DBLP:conf/dmdw/RosenthalS00,DBLP:conf/sigmod/RizviMSR04}, which are SQL \texttt{SELECT} statements---parameterized by session parameters like the current user ID---that define information in the database that a user is allowed to access.

Under a view-based policy, a query is allowed only if it can be fully answered using the views.
This criterion extends to a program that (conditionally) issues multiple queries, by treating the program as ``one big query'' that returns the results of its constituent queries.
These notions can be made precise based on \emph{query determinacy}~\cite{Nash10:determinacy},
but in the interest of space, we omit the formal definitions and offer an example instead.%
\footnote{We refer interested readers to the literature for both theoretical~\cite{lmss95,Zhang05:authorization-views,Nash10:determinacy,Afrati19:answering} and practical~\cite{DBLP:conf/sigmod/RizviMSR04,DBLP:conf/osdi/ZhangSCPSS22,DBLP:conf/sigmod/BenderKG14,DBLP:conf/sigmod/BenderKGK13} treatments of this subject.}

\begin{listing}
    \caption{A handler that displays a course's grade sheet.}\label{lst:view_grades}%
    \begin{lstlisting}[language=python, escapeinside=!!, showstringspaces=false,
                       numbers=left, numbersep=4pt,
                       aboveskip=0pt, belowskip=0pt, xleftmargin=14pt]
def view_grade_sheet(db, session, req):
  role = db.sql(!\label{q1}!
    "SELECT * FROM roles "
    "WHERE user_id = ? AND course_id = ?",
    session["user_id"], req["course_id"])
  if role is None: raise Http404!\label{line:view-grades:empty}!
  if not role.is_instructor: raise Http403!\label{line:view-grades:instructor}!
  all_grades = db.sql(!\label{q2}!
    "SELECT * FROM grades "
    "WHERE course_id = ?", role.course_id)
  return format_html(all_grades, ...)!\label{line:view-grades:html}!
    \end{lstlisting}
\end{listing}

\begin{listing}
    \caption{An example policy for the handler in \Cref{lst:view_grades}.}\label{lst:view_grades_policy}
    \begin{EV}
        \item \begin{compactsql}
SELECT * FROM roles
WHERE user_id = MyUserId
            \end{compactsql}
            \textit{A user can view their role (if any) in any course.}\label{v1}
        \item \begin{compactsql}
SELECT grades.* FROM roles, grades
WHERE roles.user_id = MyUserId
  AND roles.is_instructor
  AND grades.course_id = roles.course_id
   \end{compactsql}
     \textit{An instructor for a course can view all grades.}\label{v2}
    \end{EV}
\end{listing}

\begin{example}\label{example:running}
    Suppose a course-management site has a web request handler that displays a course's grade sheet~(\Cref{lst:view_grades}).
    It ensures that the user is an instructor before fetching grades.

    Note that the handler has access to both \emph{session parameters} (user ID) and \emph{request parameters} (course ID).
    Session parameters are trusted (e.g., set by an authentication mechanism), and may appear in the policy and dictate the extent of allowed data access.
    Request parameters are untrusted (e.g., parsed from an HTTP request) and must not appear in the policy.

    A policy for this handler might look like \Cref{lst:view_grades_policy}, where \texttt{MyUserId} denotes the user-ID session parameter.
    Notably, it does not reference the course-ID request parameter, instead allowing the handler's queries \emph{for any course ID}.
    This policy precisely captures the information the handler can query.
\end{example}

Like prior work in database access control~\cite{DBLP:conf/icde/AgrawalBGKLR05,DBLP:conf/sigmod/BenderKG14,DBLP:conf/sigmod/BenderKGK13,DBLP:conf/vldb/LeFevreAERXD04,DBLP:conf/hotos/MarzoevASYKMKM19,DBLP:conf/sigmod/RizviMSR04,shi09:soundness,DBLP:conf/osdi/ZhangSCPSS22}, we focus on extracting policies for database \emph{reads} (\texttt{SELECT}s) only.
Similar techniques can be used to extract conditions for other operations, although our policy language~(\Cref{sec:overview:scope}) and algorithms~(\Cref{sec:policy}) would need to be extended.

%% file: overview.tex
Given an application, the ideal policy extractor would produce a view-based access-control policy that satisfies:
\begin{description}
    \item[Completeness] Allows all queries the application can issue.
    \item[Tightness] Reveals as little information as possible subject to completeness.
    \item[Conciseness] Has a short representation in SQL.
\end{description}
For example, given \Cref{lst:view_grades} (but written in a real web framework), we would like to extract the policy in \Cref{lst:view_grades_policy}.

Policy extraction is challenging, and \name is not guaranteed to meet all three goals (see \Cref{sec:overview:scope}).
Nevertheless, we show in \Cref{sec:eval} that \name produces policies that are useful in practice.

\subsection{Workflow}\label{sec:overview:workflow}
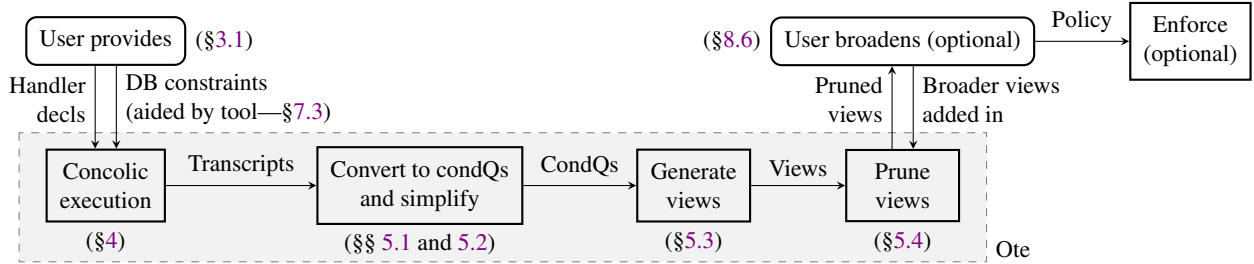
\begin{figure*}\centering
\begin{tikzpicture}[
    font=\small,
    process/.style={rectangle, text centered, align=center, inner sep=5pt, minimum width=1.5cm, draw=black, thick},
    user/.style={process, rounded corners},
    arrow/.style={->,>=stealth}
    ]
    \node (concolic) [process, label={[xshift=1.2em]below:(\Cref{sec:concolic})}] {Concolic\\execution};
    \node (convert) [process, right=1.8cm of concolic, label=below:(\Cref{sec:policy:preprocess,sec:policy:simplify})] {Convert to condQs\\and simplify};
    \node (view-gen) [process, right=1.4cm of convert, label=below:(\Cref{sec:policy:generate-sql})] {Generate\\views};
    \node (prune) [process, right=1.1cm of view-gen, label=below:(\Cref{sec:policy:prune})] {Prune\\views};

    \node (provide) [user, above=1.1cm of concolic, label=right:(\Cref{sec:overview:workflow})] {User provides};
    \node (review) [user, right=1cm of prune] {User\\reviews};
    \node (broaden) [user, above=.8cm of review, label=right:(\Cref{sec:eval:broaden})] {Broaden\\policy};
    \node (fix) [user, below=.8cm of review, label=right:(\Cref{sec:eval:compare})] {Fix app bug};
    \node (enforce) [process, right=.8cm of review] {Enforce\\(optional)};

    \draw [arrow] (concolic) -- node[above]{Transcripts} (convert);
    \draw [arrow] (convert) -- node[above]{CondQs} (view-gen);
    \draw [arrow] (view-gen) -- node[above]{Views} (prune);

    \draw [arrow] ([xshift=-4pt] provide.south) -- node[left, yshift=3pt, align=right]{Handler\\decls} ([xshift=-4pt] concolic.north);
    \draw [arrow] ([xshift=4pt] provide.south) -- node[right, yshift=3pt, align=left]{Database constraints\\(aided by a tool---\Cref{sec:implementation})} ([xshift=4pt] concolic.north);

    \draw [arrow] (prune) -- node[above]{Policy} (review);
    \draw [arrow] (review.north) -- (broaden.south);
    \draw [arrow] (review.south) -- (fix.north);
    \draw [arrow] (review) -- (enforce);
    \draw [arrow] (broaden.west) -- node[above, sloped]{Broader views} (prune.north east);
    \draw [arrow] (fix.west) -| (concolic.south);

    \begin{scope}[on background layer]
        \draw[draw=gray,dashed,fill=black!5] ([shift={(-10pt,6pt)}] concolic.north west) rectangle ([shift={(10pt,-15pt)}] prune.south east);
    \end{scope}
    \path ([shift={(-10pt,10pt)}] concolic.north west) -- node[midway, above=0pt] {\textbf{\name}} ([shift={(10pt,10pt)}] prune.north east);
\end{tikzpicture}
\caption{{\bf Policy extraction workflow.}
``CondQs'' stands for conditioned queries (\Cref{sec:policy:preprocess}).
}\label{fig:workflow}
\end{figure*}

Before discussing how we approach these goals, we first describe \name's workflow from a user's perspective~(\Cref{fig:workflow}).

Suppose a user wants to extract a policy from a web application.
We assume that the application is written in a supported framework (Ruby on Rails) and that the user is familiar with the application's functionality.

The user starts by declaring the handlers to analyze and providing the application's \emph{database constraints}.
\name supports two forms of database constraints, which cover all the constraints we encountered:
\begin{enumerate}
  \item \emph{A set of columns in a table is unique.}
  For example, a uniqueness constraint on \texttt{roles(user\_id, course\_id)} states that a user has at most one role in a course.
  \item \emph{A query $Q_1$'s result is contained in a query $Q_2$'s result.}
  Such constraints are commonly used to express foreign-key invariants: e.g., every \texttt{assignments.course\_id} must appear as some \texttt{courses.id}.
\end{enumerate}
These constraints describe the application's valid database states;
\name uses them to keep exploration within those states during concolic execution (\Cref{sec:concolic:modeling}) and to simplify the generated policy (\Cref{sec:policy:simplify}).
To reduce the user's burden, \name provides a tool that can automatically generate most required constraints from a Ruby on Rails application's schema and models (\Cref{sec:implementation,sec:eval:app-setup:constraints}).

The user then packages the application into a Docker container and invokes \name, which:
\begin{enumerate}
    \item Explores execution paths through the handlers via concolic execution, producing \emph{transcripts} that record the branches taken and the SQL queries issued (\Cref{sec:concolic});
    \item For each individual handler, analyzes the transcripts to generate a preliminary policy allowing each query to be issued under its recorded conditions (\Crefrange{sec:policy:preprocess}{sec:policy:generate-sql});
    \item Merges the individual policies and prunes any redundant views, producing a final policy (\Cref{sec:policy:prune}).\label{step:prune}
\end{enumerate}

Next, the user inspects the generated policy, using their domain knowledge of the application's privacy requirements:
\begin{itemize}
    \item A policy that is too broad can indicate an access-control bug. The user can investigate the queries that yielded a too-broad view using inputs logged by concolic execution (\Cref{sec:implementation}) and modify the application if appropriate.
    \item A policy that is too tight can be \emph{broadened} to permit more intended data accesses.  This may simplify the policy by removing filters or by replacing multiple views with a single one.
        \name assists with policy broadening: The user adds a broader view to the policy, either before or after view pruning, and then invokes pruning (Step~\ref{step:prune}) to automatically remove now-redundant views~(\Cref{sec:eval:broaden}).
\end{itemize}
Once satisfied, the user can optionally enforce the policy to ensure the application's current and future compliance.

\subsection{Assumptions, Scope, and Utility}\label{sec:overview:scope}
\paragraph{Queries}
At its core, \name supports project-select-join (PSJ) queries in set semantics.
These are queries that
\begin{inparaenum}[(1)]
\item always return distinct rows, and
\item have the form:\\
    \begin{minipage}{\linewidth}
        \begin{minipage}{.8\linewidth}
            \begin{lstlisting}[escapeinside=||]
SELECT |\textcolor{lightgray}{\bf\ttfamily [DISTINCT]}| col1, col2, ...
FROM tbl1, tbl2, ... WHERE ...
            \end{lstlisting}
        \end{minipage}\hfill
        (PSJ)
    \end{minipage}
\end{inparaenum}
Also supported are common queries that \name can mechanically rewrite into this form,
such as queries with inner joins or of the form \lstinline{SELECT 1 FROM tbl1, tbl2, ... WHERE ... LIMIT 1}.

We found that the distinct-rows assumption does not limit utility---we have never encountered queries that may return duplicate rows in our evaluation.
But applications do issue queries more complex than PSJ; this is handled differently in different stages:
The concolic-execution driver precisely models richer SQL features (\Cref{sec:concolic:modeling}),
but view generation and pruning must approximate complex queries using PSJ (\Cref{sec:policy:generate-sql,sec:eval:stats}).
We plan to extend our prototype to support more complex queries.

\paragraph{Policies}
\name generates policies consisting of PSJ views, which are expressive enough for the applications we studied.
A notable limitation is that PSJ views cannot generally express negations (e.g., ``$Q_1$ is allowed only if $Q_2$ returns no rows'').
Negations complicate approximating a query's information content~\cite[\S{}2.2]{Wang07:correctness} and can slow down enforcement~\cite[\S{}6]{DBLP:conf/osdi/ZhangSCPSS22},
so we defer handling negations to future work.

\paragraph{Non-guarantees}
\name \emph{does not guarantee} completeness or tightness:
It may generate a policy that disallows a query issued by the application, or one that can be tightened while allowing the same application queries.
This is because:
\begin{itemize}
    \item Concolic execution can miss execution paths as the input space explored is bounded.
    \item Uninstrumented operations in the \emph{query-issuing core} can lead to incomplete path conditions (\Cref{sec:concolic:simple}).
    \item A pruning optimization uses an LLM-based relevance judge, which can make mistakes (\Cref{sec:concolic:pruning}).
    \item Approximating complex queries using PSJ can make a policy more, or less, restrictive than ideal (\Cref{sec:eval:stats}).
\end{itemize}
In general, complete-and-tight policy extraction for Turing-complete code is impossible~\cite{rice}.
But in practice, \name can produce policies more accurate than handwritten ones (\Cref{sec:eval:compare}).

Similarly, \name \emph{does not guarantee} that it produces the most concise SQL representation,
but its simplification and pruning steps~(\Cref{sec:policy}) make it feasible to inspect the policy~(\Cref{sec:eval:compare}).

\paragraph{Application assumptions}
\name uses a modified Ruby interpreter and Ruby on Rails framework for concolic execution~(\Cref{sec:concolic:instrumentation}),
and so it supports only applications written in Rails.
However, our approach generalizes to other languages.

\name's effectiveness depends on a ``simple query-issuing core'' assumption (\Cref{sec:concolic:simple}).
In short, we assume that the application's SQL query issuance depends only on \emph{simple expressions}---%
ones that use only operations instrumented by \name's Ruby interpreter.
Most notably, \name does not instrument string formatting (doing so would complicate SMT solving),
and so it requires the application to issue SQL queries in parameterized form rather than splicing parameters into query strings within Ruby.
Fortunately, idiomatic Rails code already issues parameterized queries by default,
and the few places in our evaluation where this is not the case were easy to fix~(\Cref{sec:eval:app-setup}).

%% file: exploration.tex
\name begins by exploring paths through application code via concolic execution.
We open this section with some empirical observations about typical web applications that motivated our use of concolic execution.

\subsection{Observation: Simple Query-issuing Cores}\label{sec:concolic:simple}
Given a web application's codebase, consider all program statements that issue SQL queries.
Informally, imagine the backward program slice~\cite{DBLP:journals/tse/Weiser84} from these statements.
This slice, which we call the \emph{query-issuing core}, is the part relevant to policy extraction:
It consists of the program components on which query issuance has a control- or data-dependence.
Thus, it includes code that computes query parameters or determines whether later queries are issued, but excludes code used only for localization, formatting, HTML/JSON generation, logging, etc.
In \Cref{lst:view_grades}, for instance, the two queries~(Lines~\ref{q1} and~\ref{q2}) are in the query-issuing core, as are the emptiness and instructor checks~(Lines~\ref{line:view-grades:empty} and~\ref{line:view-grades:instructor}) that determine whether the second query is issued; the final HTML-rendering line~(Line~\ref{line:view-grades:html}) is outside the core.

While the codebase as a whole can be complex, it has been observed~\cite{DBLP:conf/osdi/Chlipala10,DBLP:conf/pldi/0001R19} that the query-issuing core of a typical web application is often \emph{simple}.
We confirm this observation---for the applications we studied, the core consists mostly of:
\begin{enumerate}
    \item Conditionals that test if a query's result set is empty~\cite[\S~2]{DBLP:conf/pldi/0001R19} or check basic conditions on primitive values (e.g., checking for equality or nullity)~\cite[\S~6.2]{DBLP:journals/toplas/ShenR21};
    \item Loops over a query's result set~\cite[\S~2]{DBLP:conf/pldi/0001R19} with no loop-carried dependencies~\cite[\S~4]{DBLP:conf/osdi/Chlipala10}; and
    \item Trivial data-flow to query statements---e.g., passing a value returned by one query to another.
        (Given that the application issues parameterized SQL queries, no query-string formatting operation appears in the data flow.)
\end{enumerate}

Our goal is therefore to analyze this simple query-issuing core without spending resources on the surrounding codebase.
This is challenging because the query-issuing core is not syntactically isolated.
For instance, in a typical Ruby on Rails application, SQL queries may be issued from controller actions, view templates, view helpers, and external libraries, each of which can also contain substantial code unrelated to query issuance.
To address this challenge, we use concolic execution, which allows \name to analyze the query-issuing core without syntactically isolating it.

\subsection{Concolic Execution: What and Why}\label{sec:concolic:what}
In concolic execution~\cite{DBLP:conf/pldi/GodefroidKS05,DBLP:conf/sigsoft/SenMA05},
a program is executed repeatedly using concrete inputs that have symbolic variables attached.
As the program runs, its state is tracked both concretely and symbolically.
When the program branches on a symbolic condition, the condition and its outcome are recorded.
This produces a conjunction of constraints---the \emph{path conditions}---that led execution down a path.
The conjuncts are then negated using a solver to generate new inputs (up to a bound) that will steer execution down other paths.

We chose concolic execution
because it offers a ``pay-as-you-go'' model for symbolic tracking:
We selectively instrument the operations that might appear in the query-issuing core, and the rest will simply execute concretely by default.
This strategy reduces not only our instrumentation effort, but also the number of paths explored---a branch on an uninstrumented condition will not lead to a new path.

Note that if an application's query-issuing core is \emph{not simple}---so it contains uninstrumented operations---then \name may miss certain queries or conditions, leading to a policy that is either broader or tighter than ideal.
We discuss this case, with examples and mitigations, in \Cref{sec:concolic:not-simple}.

\subsection{System Architecture}\label{sec:concolic:architecture}
\name has a \emph{driver} that generates inputs and concurrent \emph{executors} that run application code on each input.
The driver tracks explored paths in a prefix tree;
for every prefix, it negates the last condition and invokes an SMT solver to generate a new input.
(The driver keeps the prefix tree in memory and generates inputs sequentially, but both can be relaxed to improve performance.)
It then sends the input to an executor, which runs a handler using an instrumented Ruby interpreter and Ruby on Rails framework~(\Cref{sec:concolic:instrumentation}) and sends back a \emph{transcript} capturing the path conditions and queries issued (\Cref{sec:concolic:modeling}).
Exploration terminates when all prefixes have been visited.

\subsection{Symbolic Modeling and Input Generation}\label{sec:concolic:modeling}
Concolic execution requires symbolically modeling the handler's inputs, consisting of the database and session/request parameters.
To ensure termination, the input space must be bounded.
Following prior work~\cite{DBLP:conf/ccs/ChaudhuriF10}, we model the database as tables containing a bounded number of symbolic rows (our prototype uses a bound of~2).
The driver asserts the application's database constraints (\Cref{sec:overview:workflow}), so it explores only valid database states.
Inspired by UrFlow's loop analysis~\cite[\S~4.2]{DBLP:conf/osdi/Chlipala10}, we also restrict the input space so that each query returns at most one row.
For simplicity, we will describe our algorithms under this assumption, even though they can be extended to handle queries returning multiple rows.

\begin{listing*}
    \caption{A transcript from a run of the handler in \Cref{lst:view_grades}, when the user is an instructor for the course.}\label{lst:transcript}
    \begin{enumerate}
        \item $\textsc{Query}_1(
            \text{\lstinline[language=sql]{SELECT * FROM roles WHERE user_id = ? AND course_id = ?}},
            \langle\texttt{MyUserId}, \texttt{CourseId}\rangle, \textit{isEmpty}=\textit{false})$
        \item $\textsc{Branch}(r_1.\textit{is\_instructor}, \textit{outcome}=\textit{true})$
        \item $\textsc{Query}_2(
            \text{\lstinline[language=sql]{SELECT * FROM grades WHERE course_id = ?}}, \langle r_1.\textit{course\_id} \rangle, \textit{isEmpty}=\textit{false}
            )$
    \end{enumerate}
\end{listing*}

Under this modeling, a transcript is a sequence of operations performed by the application with two types of records:
\begin{enumerate}
    \item $\textsc{Query}_i(\textit{sql}, \textit{params}, \textit{isEmpty})$, meaning the $i$\textsuperscript{th} query issued was the query $\textit{sql}$ with parameters $\textit{params}$, and the result set was empty if $\textit{isEmpty}$ is true.%
    \footnote{\name assumes that the application always inspects the result set of a query by branching on its emptiness, and does not track it as a separate \textsc{Branch}.}
    If not empty, a symbol $r_i$ is introduced to represent the result row.
    \item $\textsc{Branch}(\textit{cond}, \textit{outcome})$, meaning the condition $\textit{cond}$ was branched on, and the \textit{outcome} (either true or false) branch was taken.
    The condition can reference session and request parameters as well as columns returned by previous queries (e.g., ``$r_5.\textit{author\_id} = \textit{MyUserId}$'').
\end{enumerate}
\Cref{lst:transcript} shows an example transcript from a run of the handler in \Cref{lst:view_grades}, when the user is an instructor.

Our SMT encoding represents bounded database tables using conditional tables~\cite{Imielinski84} and uses the theory of integers to model all database values~\cite{he2024verieql}, including timestamps and strings.
Each nullable value is accompanied by a boolean indicating if it is \texttt{NULL}.
This simple encoding naturally supports equality and arithmetic operations but not string operations,
a limitation that can be lifted using a string solver~\cite{DBLP:conf/fmcad/BerzishGZ17,DBLP:conf/cav/LiangRTBD14}.

Similar to prior work~\cite{he2024verieql}, we encode a subset of SQL into SMT on bounded symbolic tables.
Our encoding supports left- and inner-joins and count- and sum-aggregations.
One notable unsupported feature is ordering, which we have not needed under the assumption that each query returns at most one row.%
\footnote{Ordering cannot be ignored when a \texttt{LIMIT} clause can truncate a result set with multiple distinct rows; we encountered no such queries in our evaluation.}
We implement several optimizations for input generation: reusing Z3 AST objects, calling the solver incrementally, and caching conflicts (resulting in infeasible paths) using unsat cores~\cite{DBLP:conf/nfm/Schwartz-Narbonne15}.

\subsection{Instrumentation and Tracking}\label{sec:concolic:instrumentation}
To maintain symbolic state, we modified the JRuby interpreter~\cite{jruby} to add an optional ``symbolic expression'' field to each Ruby object.
For each class that we want represented symbolically, we implement a \texttt{with\_sym} method that returns a clone with a symbolic expression attached, and amend methods that we want instrumented to attach symbolic expressions to their results.
Unmodified methods simply return an object with no expression attached, concretizing the result as desired.

We implemented symbolic representations of ten classes (\texttt{String}, \texttt{Fixnum}, etc.) covering all symbolic inputs in our evaluation, and instrumented simple operations (equality, null-check, etc.) that appear in the query-issuing core.
We ensured that instances of a singleton object (\texttt{true}, \texttt{false}, \texttt{nil}) with different symbolic expressions are treated as equal, and that a mutating method clears an object's symbolic expression by default.

To track queries and branches,
we added a library for maintaining transcript records, exposing methods \texttt{record\_query} and \texttt{record\_branch}~(\Cref{sec:concolic:modeling}).
We modify the Rails database layer to call \texttt{record\_query} after every query, and modify JRuby's \texttt{isTrue} and \texttt{isFalse} methods (which evaluate an object's truthiness and falsiness) to call \texttt{record\_branch} with the object's symbolic expression if it has one.
The \texttt{record} methods also save the stack trace at which the query or branch is evaluated, for use in relevance-based pruning (\Cref{sec:concolic:pruning}).

\subsection{Skipping Irrelevant Branches}\label{sec:concolic:pruning}
    Concolic execution suffers from \emph{path explosion}: The number of feasible paths can grow exponentially with the number of branches.
    \name is no exception: Without further optimization, five web handlers in our evaluation (\Cref{sec:eval:stats}) failed to finish after ten hours and would likely have required days.
    Thus, we must aggressively reduce the paths explored, but without missing any query-issuing behavior.

    \paragraph{Irrelevant conditionals}
    Even with selective instrumentation (\Cref{sec:concolic:what}), many tracked branches lie outside the query-issuing core.%
    \footnote{Even if the core is simple, not every simple condition is in the core!}
    Consider the following example:

    \begin{listing}[H]
    \caption{An irrelevant conditional from Autolab.}\label{lst:irrelevant-conditional}
    \begin{lstlisting}[language=ruby, escapeinside=!!, showstringspaces=false]
<% if !\colorbox{gray!30}{submission.version == 0}! then %>
  <font size=-2>Unofficial</font>
<% else %>
  <%= submission.version %>
<% end %>
    \end{lstlisting}
    \end{listing}

    The highlighted conditional
    affects only HTML generation and has no bearing on data access;
    negating such conditionals produces new paths that reveal no new query behavior.
    Skipping these \emph{irrelevant} conditionals can reduce the exploration space dramatically, since each may cut the path count in half.

    \begin{listing}
    \caption{
        An abridged relevance-judge output for the conditional in \Cref{lst:irrelevant-conditional}, produced by \texttt{gpt-5} in 24~seconds (\Cref{sec:eval:relevance-judge}).
    }\label{lst:irrelevant-conditional-output}
    \begin{lstlisting}[language={}, showstringspaces=false, breaklines=true, basicstyle=\ttfamily\footnotesize]
IRRELEVANT

The conditional `submission.version == 0` [...] only controls presentation text:
- If true, it renders the literal "Unofficial".
- If false, it renders the numeric `submission.version`.
[...]
    \end{lstlisting}
    \end{listing}

    \paragraph{Identifying irrelevance using an LLM}
    Statically identifying irrelevant conditionals is hard in Ruby on Rails~\cite{Smit2025} due to dynamic language features and runtime template compilation~\cite[\S~6.2]{PaoloPerrotta2014MR2}.
    However, many irrelevant branches follow recognizable patterns---e.g., conditionals that issue no queries and mutate no state.
    Rather than manually encoding such patterns as heuristics, we draw on recent successes in using LLMs to complement software analysis~\cite{DBLP:journals/pacmpl/LiHZQ24,DBLP:conf/sosp/StoicaSSZ0MMN24,DBLP:conf/llm4code/MunsonGC25,DBLP:journals/corr/abs-2504-17542} and delegate the identification of irrelevant conditionals to a \emph{relevance judge} implemented using an LLM-based coding agent.

    When \name encounters a conditional, it creates a prompt containing a definition of irrelevance, the conditional, and the stack trace at which the conditional was evaluated.
    It sends the prompt to a coding agent, which autonomously inspects source files and answers ``relevant'', ``irrelevant'', or ``unsure'' (treated as relevant), plus an explanation.
    The prompt is generic and contains no framework- or application-specific details.
    \name records the verdict and explanation for later inspection;
    see \Cref{lst:irrelevant-conditional-output} for an example.

    Because the relevance judge can take minutes to render a verdict (\Cref{sec:eval:relevance-judge}), \name calls it asynchronously.
    Exploration carries on, assuming the conditional is relevant;
    if the judge later designates a conditional as irrelevant, the driver retroactively records it as such and takes it into account in the future.

    \name avoids calling the relevance judge when
    \begin{inparaenum}[(1)]
        \item the conditional's designation is known (e.g., ``$Q$ returns non-empty'' is relevant if $Q$'s result is used in another query), or
        \item the conditional is \emph{vacuous}---i.e., its negation is infeasible given the relevant conditionals that precede it.
    \end{inparaenum}
    Because a conditional's vacuousness may depend on the relevance of its predecessors,
    when \name marks a conditional as irrelevant, it recomputes the vacuousness of any formerly-vacuous later conditionals.

    \paragraph{Pruning}
    Using the judge's verdicts, the driver
    avoids negating irrelevant conjuncts in path conditions,
    skips paths whose subsequence of \emph{relevant} conditions is already covered, and
    drops irrelevant conditions from the final transcript.
    
    \paragraph{Accuracy}
    Because the relevance judge uses an LLM, it cannot guarantee accuracy.
    To defend against mistakes, 
    \name exposes the verdicts and explanations for human review, and allows the human to guide the judge by adding inline comments prefixed with ``\texttt{RELEVANCE-HINT}'', which the LLM agent is instructed to consider.
    In our evaluation (\Cref{sec:eval:app-setup}), we added hints
    primarily to compensate for the coding agent's limited understanding of Rails and external libraries.
    With these hints, we verified the judge's irrelevance verdicts to be accurate.

\subsection{When the Core Is Not Simple}\label{sec:concolic:not-simple}
\input{discussion}

%% file: discussion.tex
The simple query-issuing core assumption~(\Cref{sec:concolic:simple}) is not true for all practical applications.
It can be violated in two ways.

\paragraph{Complex control-flow conditions}
This case arises when a query's issuance depends on a complex expression.
For example, \diaspora crashes if a photo's \texttt{url} is not a valid URL, thus preventing later queries;
the URL-validity condition relies on regex-matching operations, which \name does not instrument.

Suppose a query~$Q$ is issued conditioned on an expression~$C$ with some uninstrumented operations.
This may cause \name to generate a policy that is either broader or tighter than ideal:
\begin{enumerate}
  \item If \name happens to generate an input that satisfies~$C$
    (this is not guaranteed because the driver is unaware of~$C$),
    then $Q$ will appear as a conditioned query \emph{without}~$C$ in its condition,
    leading to a policy broader than ideal.\label{item:vio1}
  \item If \name never generates an input that satisfies~$C$,
    then $Q$ may not appear in any transcript, leading to a policy that may not allow~$Q$---which is tighter than ideal.%
    \label{item:vio2}
\end{enumerate}
In practice, Case~\ref{item:vio1} has not been a problem for us because
\begin{inparaenum}[(1)]
  \item the complex condition~$C$ typically reflects business logic and has no privacy implications~(\Cref{sec:eval:broaden}), and
  \item query~$Q$ is likely also issued elsewhere under a simpler condition.
\end{inparaenum}
Case~\ref{item:vio2} is problematic for validations (e.g., URL validity);
we handle this by manually setting the input using database constraints (e.g., setting URLs to always be \texttt{http://foo.com}).

If an uninstrumented operation turns out important for an application domain,
we would update \name's Ruby instrumentation, SMT encoding, and SQL generation to support it.

\paragraph{Complex data flow}
This case arises when the transcript contains an expression derived from a symbolic value through uninstrumented operations---%
a problematic scenario because the symbolic value is concretized.
For example, the conditional \lstinline[language=ruby]{if uid < uid * uid} may insert a \textsc{Branch} record for $\textit{uid}<16$ into the transcript,
as multiplication is not instrumented.
In such cases, the driver might fail to terminate because a ``new path condition'' emerges for every value of \texttt{uid}.
Even if we cut off the exploration, the generated policy would be overly strict and verbose, littered with concrete filters like $\textit{uid}<16$.

To detect such harmful concretizations, the driver reports each new constant and new query it encounters (there should be a bounded number of these).
This mechanism alerted us to a few non-parameterized SQL queries~(\Cref{sec:eval:app-setup}), where symbolic values were formatted into the query string.
We encountered no concretization issues in our final evaluation, and we plan to implement more precise detection of such ``partially concrete'' conditions via heavier-weight data-flow tracking.

%% file: policy.tex
After exploration, \name merges and simplifies the transcripts and creates a preliminary set of views for each handler.
It then gathers the views for all handlers explored and removes redundancy by leveraging an existing enforcement tool, Blockaid~\cite{DBLP:conf/osdi/ZhangSCPSS22}.
We now delve into this policy-generation process.

\subsection{Preprocessing Into Conditioned Queries}\label{sec:policy:preprocess}
As a first step, \name processes the transcripts for each handler into a set of \emph{conditioned queries}.
A conditioned query is a tuple $\tupL \textit{sql}, \textit{params}, \textit{conditions} \tupR$, where \textit{conditions} is a list of prior \textsc{Query} and \textsc{Branch} records.%
\footnote{A prior \textsc{Query} represents the \emph{condition} that a query~$Q$ returned empty or not;
the \emph{action} of $Q$'s issuance is captured by $Q$'s own conditioned queries.}
It associates each query with the conditions under which it is issued; one conditioned query is generated for each query issued in each transcript.
As \name does not currently support negations in policies (\Cref{sec:overview:scope}), it drops from
\textit{conditions} any \textsc{Query} record with $\textit{isEmpty}=\textit{true}$ (i.e., a condition that a prior query returns empty).

\subsection{Simplifying Conditioned Queries}\label{sec:policy:simplify}
For each handler, \name simplifies the set of conditioned queries by removing redundancy.
At a high level, the simplifications remove conditions that do not affect whether a query is issued.

For example, suppose two conditioned queries have the same SQL and parameters, but one is guarded by $C \land b$ and the other by $C \land \lnot b$.
Since the query is issued regardless of the branch outcome, \name replaces the pair with a single conditioned query guarded only by~$C$.
Similarly, a conditioned query guarded by $C$ subsumes one guarded by $C \land d$ with the same SQL and parameters.
Database constraints expose another source of redundancy, allowing \name to remove conditions that are guaranteed to hold, such as a prior lookup that must succeed because of a foreign-key dependency.

\name performs these and other simplifications by following the steps shown in \Cref{alg:simplify}:
\begin{itemize}
    \item Remove \textsc{Branch}es that \emph{must} be taken due to an input constraint or a prior condition~(Line~\ref{alg:simplify:remove-vacuous-br});
    \item Unify variables that are constrained to be equal by query filters
    (Line~\ref{alg:simplify:propagate-eq});
    \item Remove identical \textsc{Query} records from each conditioned query's \textit{conditions} (Line~\ref{alg:simplify:remove-dup});
    \item Drop \emph{vacuous} \textsc{Query} records---for queries that \emph{must} return a row due to, e.g., a foreign-key dependency---whose result is not subsequently referenced~(Line~\ref{alg:simplify:remove-vacuous-qr});
    \item Merge pairs of conditioned queries that differ only in the outcome of a single \textsc{Branch} record~(Line~\ref{alg:simplify:merge-branches})---the query is issued no matter which way the branch goes;
    \item Remove a conditioned query if another exists with the same \textit{sql} and \textit{params} but only a subset of the \textit{conditions}~(Line~\ref{alg:simplify:remove-subsumed})---the latter subsumes the former.
\end{itemize}
Each step is parallelized across cores.
Due to the large number of conditioned queries, we designed these steps to favor efficiency over optimality.
Any missed opportunities for simplification will be caught by the final pruning step (\Cref{sec:policy:prune}).

\begin{algorithm}
\caption{Simplifying a set of conditioned queries (\Cref{sec:policy:simplify}).}\label{alg:simplify}%
\begin{algorithmic}[1]
\ForAll{conditioned query}
    \State remove vacuous branches\label{alg:simplify:remove-vacuous-br}
    \State propagate equalities\label{alg:simplify:propagate-eq}
    \State remove duplicate queries\label{alg:simplify:remove-dup}
\EndFor
\Repeat
    \ForAll{conditioned query}\label{alg:simplify:vacuous-loop}
        \State remove vacuous-and-unused query records\label{alg:simplify:remove-vacuous-qr}
    \EndFor
    \State \algorithmicrepeat~merge branches \algorithmicuntil~\textbf{convergence}\label{alg:simplify:merge-branches}
\Until{\textbf{convergence}}\label{alg:simplify:outer-loop}
\State remove subsumed\label{alg:simplify:remove-subsumed}
\end{algorithmic}
\end{algorithm}

\subsection{Generating SQL View Definitions}\label{sec:policy:generate-sql}
\name now generates one SQL view per conditioned query.
The view should reveal the same information as the conditioned query's path if its conditions are met, and no information otherwise.
For this, \name uses an iterative algorithm that ``conjoins'' each condition record onto an accumulated query definition~$A$.
It maintains the invariant that query~$A$:
\begin{itemize}
    \item Returns empty if any previous condition is violated;
    \item Returns the Cartesian product of all prior queries' results (i.e., concatenations of one-row-per-query) otherwise.
\end{itemize}
Query~$A$ serves two purposes: It captures the branching conditions, and it exposes query results to be referenced by later records.
Lastly, the algorithm conjoins the final query onto~$A$.

Before fully specifying the view-generation algorithm, let us walk through an example.

\begin{example}\label{example:view-generation}
    Consider \Cref{lst:transcript}.
    We shall generate the view for the conditioned query associated with $\textsc{Query}_2$.

    The algorithm starts with the query~$A$ that returns the empty tuple.
    After the first record ($\textsc{Query}_1$), $A$ is updated to $A_1$:
    \begin{lstlisting}
SELECT * FROM roles
WHERE user_id = MyUserId
  AND course_id = CourseId
    \end{lstlisting}
    (We use CamelCase for request and session parameters.)

    After the second record (the \textsc{Branch}), $A$ is updated to $A_2$:
    \begin{lstlisting}
SELECT * FROM roles
WHERE user_id = MyUserId
  AND course_id = CourseId
  AND is_instructor
    \end{lstlisting}
    Observe that $A$ indeed returns the same rows as $\textsc{Query}_1$ if the branch condition holds, and an empty result otherwise.

    Then, $\textsc{Query}_2$ is conjoined onto~$A$, resulting in view~$V$:
    \begin{lstlisting}
SELECT roles.*, grades.* FROM roles, grades
WHERE roles.user_id = MyUserId
  AND roles.course_id = CourseId
  AND roles.is_instructor
  AND grades.course_id = roles.course_id
    \end{lstlisting}
    Note that $\textsc{Query}_2$'s sole parameter, $r_1.\textit{course\_id}$, has been replaced by the \texttt{course\_id} column exposed by~$A_2$.
    A final step remains to remove the \texttt{CourseId} parameter, which we will discuss at the end of this subsection.
\end{example}

A generated view can be thought of as a natural generalization of the conditioned query to cases where a SQL query can return multiple rows:
The view allows a query to be issued for every combination of rows returned by the prior queries, as long as the branching conditions are met.
This would allow the program to have loops over result sets in the form we assume in the simple query-issuing core~(\Cref{sec:concolic:simple}).

\begin{algorithm}
\caption{View generation from conditioned query (\Cref{sec:policy:generate-sql}).}\label{alg:generate-sql-view}%
\begin{algorithmic}[1]
\Procedure{GenerateSqlView}{\textit{cq}}
\State $A \gets \{\tupL\tupR\}$ \Comment{Constant query returning empty tuple}
\State $\mathcal{M}\gets \{\}$ \Comment{Maps query result column to $A$'s column}
\ForAll{$\textit{cond} \in \textit{cq}.\textit{conds}$}
    \If{$\textit{cond}$ \textbf{is} $\textsc{Branch}(\theta, \textit{outcome}=\textit{true})$}
        \State $A \gets \sigma_{\theta[\mathcal{M}]}(A)$
    \ElsIf{$\textit{cond}$ \textbf{is} $\textsc{Branch}(\theta, \textit{outcome}=\textit{false})$}
        \State $A \gets \sigma_{\lnot \theta[\mathcal{M}]}(A)$
    \ElsIf{$\textit{cond}$ \textbf{is} $\textsc{Query}_{\ell}(\textit{sql}, \textit{params})$}
        \State $Q_{\ell} \gets \textsc{SqlToRa}(\textit{sql}, \textit{params})$
        \State \Comment{Converts SQL query to relational algebra}
        \State $(A,\mathcal{M})\gets \textsc{ConjoinQuery}(Q_{\ell}, A, \mathcal{M})$
    \EndIf
\EndFor

\State $Q_{\ell+1}\gets \textsc{SqlToRa}(\textit{cq}.\textit{sql}, \textit{cq}.\textit{params})$
\State $(A,\mathcal{M})\gets \textsc{ConjoinQuery}(Q_{\ell+1}, A, \mathcal{M})$
\State \textbf{return} $\textsc{RaToSql}(A)$
\EndProcedure

\Procedure{ConjoinQuery}{$Q_{\ell}, A, \mathcal{M}$}
    \State let $Q_{\ell}=\pi_{j_1,\ldots,j_m} \sigma_{\theta} (S_1\times S_2\times \cdots)$\Comment{Normal form}
    \State $n\gets \arity(A)$ \Comment{Number of columns in $A$}
    \State $\theta'\gets \theta[k\mapsto k+n][\mathcal{M}]$ \Comment{$\forall$~column index $k$}
    \State $A\gets \pi_{1,\ldots,n,n+j_1,\ldots,n+j_m}\sigma_{\theta'}(A\times S_1\times S_2\times \cdots)$
    \State $\mathcal{M}\gets \mathcal{M}\cup\{ r_{\ell}.i \mapsto i + n \mid 1\leq i\leq m \}$
    \State \textbf{return} $(A, \mathcal{M})$
\EndProcedure
\end{algorithmic}
\end{algorithm}

We specify the view-generation procedure in \Cref{alg:generate-sql-view}, which uses relational algebra notation (under the unnamed perspective)~\cite[\S~3.2]{DBLP:books/aw/AbiteboulHV95} for conciseness.
The algorithm works for PSJ queries in normal form~\cite[Prop.~4.4.2]{DBLP:books/aw/AbiteboulHV95}.
It does not currently handle general joins or aggregations; when these arise, we approximate them using supported constructs~(\Cref{sec:eval:stats}).

\Cref{alg:generate-sql-view} essentially follows the steps illustrated by \Cref{example:view-generation}.
It maintains a mapping $\mathcal{M}$ from result columns of prior queries to columns in $A$ (e.g., mapping $r_1.\textit{course\_id}$ to the third column of $A$).
It then uses this mapping to resolve references to results from prior queries.

\paragraph{Removing request parameters}
Recall from \Cref{example:running} that request parameters like \texttt{CourseId} must not appear in view definitions.
So strictly speaking, \Cref{example:view-generation} has produced not one view, but a set of views:%
\footnote{We let $V[\texttt{X} \mapsto x]$ denote the view definition $V$ with each occurrence of \texttt{X} replaced by~$x$,
and $\dom(\texttt{col})$ denote the set of valid values for column \texttt{col}.}
\[
\left\{ V[\texttt{CourseId} \mapsto x] : x \in \dom(\texttt{roles.course\_id}) \right\}
\]
To collapse this set into one view, we use the following fact.

\begin{fact}[Informal]
    Let $V[\texttt{X}]$ be
    a view definition of the form:
    \begin{lstlisting}[escapeinside=||]
SELECT col|$_1$|, col|$_2$|, ... FROM tbl|$_1$|, tbl|$_2$|, ...
WHERE col|$_j$| = X AND |$f$|
    \end{lstlisting}
    where \texttt{col$_j$} is a non-nullable column, \texttt{X} is a request parameter, and $f$ does not refer to \texttt{X}.
    Then the set of views $\{ V[\texttt{X} \mapsto x] : x \in \dom(\texttt{col}_j) \}$
    reveals the same information as the view:
    \begin{lstlisting}[escapeinside=||]
SELECT col|$_j$|, col|$_1$|, col|$_2$|, ...
FROM tbl|$_1$|, tbl|$_2$|, ... WHERE |$f$|
    \end{lstlisting}
\end{fact}

While this fact applies only to queries of a specific form, it already covers all cases encountered in our evaluation.
We leave a theoretical study of the general case to future work.

\begin{example}[continues=example:view-generation]
Starting from view~$V$, \name removes the condition \texttt{roles.course\_id = CourseId}, but refrains from adding \texttt{roles.course\_id} to the \texttt{SELECT} statement because it is redundant with the already-present \texttt{grades.course\_id}.
This brings us to the final view $V^{\star}$:
\begin{lstlisting}[frame=single,aboveskip=5pt,belowskip=5pt]
SELECT * FROM roles, grades
WHERE roles.user_id = MyUserId
  AND roles.is_instructor
  AND grades.course_id = roles.course_id
\end{lstlisting}

The view~$V^{\star}$ differs from the handwritten view~\ref{v2} only by retaining extra columns.
However, conditioned on the existence of \ref{v1} (which \name would have generated from another conditioned query),
$V^{\star}$ reveals the same information as \ref{v2} and so is equally tight.
For simplicity, \name outputs $V^{\star}$ without removing extraneous columns,
since we find $V^{\star}$ just as concise and readable as \ref{v2}.
\end{example}

\paragraph{Outputting SQL}
\name outputs SQL views in the form:
\begin{lstlisting}
SELECT ... FROM tbl1, tbl2, ... WHERE ...
\end{lstlisting}
similar to $V^{\star}$ above.
It uses standard query-optimization passes to make the query shorter and easier to read.

\subsection{Pruning Views via Enforcement}\label{sec:policy:prune}
Lastly, \name minimizes the set of views for each handler---producing a subset that reveals the same information---and then takes their union and minimizes it again.
To minimize a set~$\mathbf{V}$ of views, \name goes through each view $V\in\mathbf{V}$ and checks whether the information revealed by~$V$ is already contained in that revealed by $\mathbf{V}\setminus\{V\}$; if so, it removes $V$.
Heuristically, \name sorts the views by the number of joins in decreasing order, so that longer views have a chance of being removed first.

It remains to check information containment.
This is the same problem as policy enforcement: checking whether issuing $V$ \emph{as a query} is allowed under the policy $\mathbf{V}\setminus\{V\}$.
To tackle this, we repurpose an existing enforcement tool, Blockaid~\cite{DBLP:conf/osdi/ZhangSCPSS22},
which we extended with a command-line interface to be invoked by \name.
Finally, \name outputs the minimized set of views for human review.

%% file: implementation.tex
\paragraph{Driver and Policy Generator}
\name's concolic-execution driver and policy generator are implemented in Scala~3.
The driver uses Apache Calcite~\cite{begoli18:calcite} to convert SQL queries into relational algebra,
and invokes Z3 using its Java binding~\cite{deMoura12:z3-java}.
It communicates with executors using Protobuf messages, and saves program inputs and transcripts to compressed JSON files.
The policy generator uses Calcite's query analyses and optimizations for conditioned-query simplification~(\Cref{sec:policy:simplify}), and uses Scala's parallel collections~\cite{scala-parallel-collections} for parallelization.

For the relevance judge~(\Cref{sec:concolic:pruning}), \name invokes parallel instances of the Codex CLI~\cite{CodexCLI} coding agent using the non-interactive \texttt{exec} subcommand.
\name parses the agent's textual output for the verdict and logs the details for human review, treating any ``unsure'' verdict or malformed output as relevant.

\paragraph{Executors}
\name's library uses RSpec~\cite{rspec} to invoke handlers (``controller actions'') with inputs from the driver.
At startup, the executor clears the database.
For each input, it begins a database transaction, populates the database, installs symbolic request parameters by patching the \texttt{params} hash,
and sets two symbolic session parameters:
the user ID and the current time.
It then invokes the handler and rolls back the transaction.

We disable Rails's fragment and low-level caching to expose queries issued only on cache misses.
As an optimization, we also disable Rails's query cache, which introduces branches that do not affect what data is fetched.
We execute handlers on MySQL backed by an in-memory \texttt{tmpfs}, and configure string columns to use a case-sensitive collation~\cite{mysql-case-sensitive}
as our SMT encoding (\Cref{sec:concolic:modeling}) does not support case insensitivity.

\paragraph{Generating database constraints}
To help the user write down an application's database constraints (\Cref{sec:overview:workflow}),
\name provides a tool that generates common types of constraints for a Rails application,
by inspecting the database schema's SQL constraints and the Active Record models' validators, associations, and inheritance hierarchy.
The user may then supplement the list with constraints not covered by the tool (\Cref{sec:eval:app-setup}), a task we hope to automate using LLMs in the future.

\paragraph{Tracing a view back to the application}
To help users understand why a view was generated, \name records with each view the ID of an execution from which the view is derived.
Using this ID, the user can recover the corresponding input and re-run the execution---optionally under a debugger---to inspect the program state when a query was issued.

%% file: eval.tex
\input{figures/stats-macros}
We applied \name to three existing applications:
\begin{enumerate}
  \item \diaspora~\cite{diaspora}, a social network with over 850~k users;
  \item Autolab~\cite{autolab}, a platform for managing course assignments used at over 20~schools; and
  \item The Odin Project (Odin)~\cite{theodinproject}, a site where over a million users take web development classes and share their work.
\end{enumerate}
We chose open-source Ruby on Rails applications with nontrivial access-control logic, while staying within the scope of the \name prototype.
Within this space, \diaspora and Autolab were natural choices because we had written policies for them by hand before~\cite[\S~8.1]{DBLP:conf/osdi/ZhangSCPSS22},
allowing us to compare those policies with the extracted ones.
Odin provides a complementary case:
We had never worked with it before, so our experience applying \name to Odin was not biased by prior knowledge.

The key takeaways from this evaluation are:
\begin{itemize}
  \item \name can extract a policy within a few hours (\Cref{sec:eval:perf}).
  \item The extracted policies avoided several errors present in the handwritten policies and revealed an access-check bug in the application code (\Cref{sec:eval:compare}).
  \item Applying \name required limited manual setup and review effort (which we summarize in \Cref{sec:eval:manual-effort}).
\end{itemize}

\begin{table}
    \caption{{\bf Number of database constraints} (\Cref{sec:eval:app-setup:constraints}).
    }\label{tab:constraints}
    \centering\small\ra{1.1}%
    \begin{tabular}{
        l 
        r 
        r 
        r 
        r 
    }
    \toprule
    & \multicolumn{2}{c}{\bf Application Logic} & {\bf For \name} & \\
    \cmidrule(lr){2-3} \cmidrule(lr){4-4}
    & Generated & Manual & Manual & {\bf Total} \\
    \midrule
      {\bf \diaspora}   & \num{347} & \num{63} & \num{2} & \num{412} \\
      {\bf Autolab}     & \num{185} & \num{42} &  \num{4} & \num{231} \\
      {\bf Odin}        & \num{116} &  \num{9} &  \num{1} & \num{126} \\
      \bottomrule
    \end{tabular}
\end{table}

\subsection{Application Setup}\label{sec:eval:app-setup}
All applications run on MySQL in our experiments.
Their database schemas are nontrivial:
\diaspora has \num{50}~tables and \num{387}~total columns,
Autolab has \num{26}~tables and \num{269}~total columns,
and Odin has \num{17}~tables and \num{152}~total columns.

\subsubsection{Code changes}\label{sec:eval:app-setup:code}
We used versions of \diaspora~(v0.7.14) and Autolab~(v2.7.0) that had previously been modified to work with Blockaid.
To summarize the most relevant modifications:
\begin{itemize}
  \item We had modified a few code locations to issue parameterized queries.
    (Most queries were already parameterized, but some needed a rewrite---see below for an example.)
  \item We had modified the applications to fetch sensitive data only if it affects output.
    (These modifications allow for finer-grained policies but are \emph{not required by \name};
    we kept them for an apples-to-apples policy comparison.)
  \item We had rewritten one query to an equivalent form supported by \name.
    (This modification is \emph{not fundamentally required}---it could have been implemented in \name itself.)
\end{itemize}
In addition, we modified seven lines in Autolab to save query results in variables for reuse.
This simple refactor helped the relevance judge identify more irrelevant conditionals by clarifying that accessing the variables does not trigger a query.

For Odin, we made three changes (over commit \texttt{f6762f0}):
\begin{itemize}
  \item We modified two locations to issue parameterized SQL queries:
    e.g., from
    \lstinline[language=ruby]!where('expires >= ?', Time.now)!
    to
    \lstinline[language=ruby]!where(expires: Time.now..)!,
    because the former inserts \texttt{Time.now} into the query string within Ruby, resulting in non-parameterized SQL.
  \item We deleted one line to prevent an aggregation from appearing in the \textit{conditions} portion of conditioned queries (currently unsupported by \name); this change preserved application behavior.
\end{itemize}

\begin{table*}[t]
    \caption{{\bf Statistics and performance.} ``\prune{}'' marks runs that used the relevance judge~(\Cref{sec:concolic:pruning}).
    Under \underline{Statistics},
    ``\#Cond.~Queries'' shows the number of conditioned queries before and after simplification~(\Cref{sec:policy:simplify});
    ``\#SQL Views'' shows the number of views after per- and cross-handler pruning~(\Cref{sec:policy:prune}).
    Under \underline{Running Time}, ``Simplify'' stands for conditioned-query simplification and view generation~(\Cref{sec:policy:simplify,sec:policy:generate-sql}),
    ``Prune'' for per-handler view-pruning, ``Final Prune'' for cross-handler view-pruning~(\Cref{sec:policy:prune}), and ``Total'' for the total sequential running time.
    }\label{tbl:main}
    \centering%
    \resizebox{\textwidth}{!}{
      \input{figures/main-table}
    }
\end{table*}

\subsubsection{Relevance hints}\label{sec:eval:app-setup:hints}
We added seven \texttt{RELEVANCE-HINT} comments (\Cref{sec:concolic:pruning}):
one to \diaspora and six to Autolab.
(Odin had few enough branches that the relevance-judge optimization was not needed---see \Cref{tbl:main}.)
Two of the hints clarify how an external library affects queries;
three note that an already-loaded expression does not trigger a query;
one notes that an expression \emph{may} issue a query;
and one encodes the assumption that an Autolab scoring function issues no queries.

All of these hints, except the last, compensate for limitations in the LLM-based coding agent's reasoning.
We expect these hints to become unnecessary with improvements in the agent's capabilities, or in how it is used:
e.g., providing the agent with library documentation,
details on Rails's ORM,
and runtime information about where queries originate.

\subsubsection{Database constraints}\label{sec:eval:app-setup:constraints}
\Cref{tab:constraints} summarizes the database constraints supplied to \name.
Over \SI{80}{\percent} were auto-generated (\Cref{sec:implementation}).
The rest were added manually: Some were transcribed from application logic, while those in the ``For \name'' column were introduced to scope the extracted policy or reduce exploration time.

Of the manually written application-logic constraints, many captured application-level invariants---%
e.g., two \diaspora posts cannot be reshares of each other---%
and could not be auto-generated.
For fields with constraints either unsupported by our SMT encoding (e.g., URL validity) or related to the external environment (e.g., an Autolab course name must correspond to a directory on disk),
we constrained them to a fixed set of valid values after creating any required external state.

We wrote seven additional constraints specifically for policy extraction.
To save exploration time, we required nullable string fields to be non-empty (as the two cases are typically treated identically), limited the Autolab analysis to non-admin users (as admin policies are trivial to write), and fixed the \diaspora user's language to English.
In Autolab, we disabled scheduler actions (which are not governed by user-specific policies)
and disallowed zero score-penalties because our prototype does not track float operations
(this preserves application features as a \texttt{0.0} penalty is equivalent to no penalty).

\subsection{Experiment Setup}\label{sec:eval:experiment-setup}
We ran experiments on Google Compute Engine using a \texttt{c3-standard-176} instance.
The driver launched 48~parallel executors
and used Z3~v4.11.2 for SMT solving.
It invoked Codex CLI (v0.58.0) using the \texttt{gpt-5} model and medium reasoning effort, with a maximum parallelism of~16, a timeout of \SI{5}{\minute}, and at most three retries after timeout.
For view pruning, it invoked Blockaid with a timeout of \SI{5}{\second}.
Executors ran a modified version of JRuby~v9.3.13.0 atop OpenJDK~21.

We ran path exploration on each handler three times; \Cref{tbl:main} shows the data from the runs with median exploration time.
To save on LLM calls, we enabled relevance-based pruning (\Cref{sec:concolic:pruning}) only for those handlers whose unpruned exploration exceeded \SI{15}{\minute} (marked with \prune{} in \Cref{tbl:main}).

\subsection{Paths, Conditioned Queries, and Views}\label{sec:eval:stats}
\Cref{tbl:main} shows that while the number of explored paths can be large,
\name is able to reduce the number of views to between \theodinprojectFinalViews and \autolabFinalViews---%
a manageable number for human review.

For view generation, we had to approximate some SQL features unsupported by either our \name prototype or Blockaid:
\begin{itemize}
  \item We rewrite a \texttt{LEFT JOIN} into an equivalent \texttt{INNER JOIN} if possible.
    Otherwise, we split it into an \texttt{INNER JOIN} (for rows that match) and a \texttt{SELECT} (for rows that don't).
  \item We rewrite a \lstinline[language=sql]{SELECT COUNT(*) FROM tbl} in the query portion into \lstinline[language=sql]{SELECT id FROM tbl},
    and omit aggregations in the condition portion.%
    \footnote{The latter did not end up affecting the final policies because every aggregation present in the condition was redundant.}
  \item We omit date-timestamp comparisons and replace ``\texttt{Now + 1 second}'' with ``\texttt{Now}''.
\end{itemize}
These approximations are ``lossy'': They can broaden or tighten the policy and must be applied with human discretion.

For Autolab, we applied a simple broadening immediately after view generation (\Cref{fig:workflow}).
The \texttt{Assessments\#show} handler queries the \texttt{score\_adjustments} table under many complex conditions, producing a large number of views.
Since the table contains only non-sensitive metadata, we added a single broad view:
\lstinline[language=sql]{SELECT * FROM score_adjustments}, which caused view pruning to eliminate all narrower views involving the table.
\Cref{tbl:main} reports results with this broadening applied.
We give a more involved example of broadening in \Cref{sec:eval:broaden}.

\subsection{Performance}\label{sec:eval:perf}
\Cref{tbl:main}~(right) shows the running times for each phase of policy extraction.
Overall, end-to-end extraction completes within five hours at most.
When we ran path exploration on the handlers marked with \prune{} \emph{without} the relevance-judge optimization (\Cref{sec:concolic:pruning}), all timed out after ten hours (not shown in the table) and, by our estimate, would have required \emph{many days} to finish.
This demonstrates that relevance-based pruning is essential for exploring real web-application code at scale.

\begin{table}
    \caption{\textbf{Relevance-judge verdicts and running times.}
      ``Rel.'' denotes relevant; ``Irrel.'' denotes irrelevant (\Cref{sec:eval:relevance-judge}).
    }\label{tbl:relevance}
    \centering\small\ra{1.1}%
    \input{figures/relevance-table}
\end{table}

\subsection{Calls to the Relevance Judge}\label{sec:eval:relevance-judge}
\Cref{tbl:relevance} reports relevance-judge verdicts (it never returned ``unsure'') and call-duration statistics.
We manually reviewed every irrelevant verdict to confirm its accuracy.
Although there were over \num{600} irrelevant verdicts, they arose from a limited set of code locations (e.g., in \diaspora, a profile-picture helper invoked along many paths),
making grouped auditing more straightforward.
We note that the call durations include LLM reasoning time and serving latency, both outside our control.

\subsection{Findings From the Extracted Policies}\label{sec:eval:compare}
We now compare the extracted policies with the handwritten \diaspora and Autolab policies from our prior work~\cite[\S~A]{DBLP:conf/osdi/ZhangSCPSS22}.
For Odin, where no handwritten policy was available, we manually inspected the extracted policy and found it accurate for the handlers analyzed.

\begin{table}
  \caption{
    \textbf{View count in extracted vs handwritten policies.}
    Extracted policies contain more views because they often preserve conditions that handwritten policies relax~(\Cref{sec:eval:compare}).
  }\label{tbl:compare}
  \centering\small\ra{1.1}%
  \begin{tabular}{lrrr}
    \toprule
    & \textbf{\diaspora} & \textbf{Autolab} & \textbf{Odin} \\
    \midrule
    \textbf{Extracted policy} & \diasporaFinalViews & \autolabFinalViews & \theodinprojectFinalViews \\
    \textbf{Handwritten policy} & 66 & 37 & -- \\
    \bottomrule
  \end{tabular}
\end{table}

\paragraph{Expectations}
We expect extracted policies to be tighter (more restrictive) than handwritten ones:
\name aims to produce the tightest possible policy, whereas humans often relax non-privacy-critical conditions and allow accesses beyond what the application requires (\Cref{sec:eval:broaden}).
Thus, extracted policies are generally longer because they encode conditions omitted from handwritten policies~(\Cref{tbl:compare}).
In our evaluation, this extra detail remained manageable for human review.

\paragraph{Where handwritten policies reveal too much}
But not all relaxations by the human policy-writer are benign.
When we compared the extracted Autolab policy against the handwritten one, we found that the latter granted course assistants access to five types of records in a ``disabled'' course,
when the application's logic states that only \emph{instructors} should have access.
Such an erroneous policy, if enforced, would allow a future code change to leak sensitive data to course assistants.

\paragraph{Where handwritten policies reveal too little}
Unexpectedly, there is also information revealed by the extracted policy but not by the handwritten one.
To investigate, we used Blockaid to check whether each extracted view is allowed as a query under the handwritten policy.
We found that:
\begin{itemize}
  \item The handwritten \diaspora policy failed to reveal the pod of a ``remote'' person and data for \texttt{MentionedInPost} and \texttt{MentionedInComment} notifications.
  \item The handwritten Autolab policy overlooked granting instructors access to all attachments in their courses.
\end{itemize}
Enforcing a policy with such oversights would disrupt application functionality by
denying legitimate data accesses.

\paragraph{A defective access check}
While reviewing the extracted Autolab policy, we noticed that none of the submissions-related views checked the \texttt{assessments.exam} column.
This was suspicious because Autolab is supposed to prohibit students from downloading prior exam submissions.

It turned out that we had introduced a bug years earlier when adapting Autolab for access control.
When configuring the \texttt{lazy\_column} gem~\cite{lazycolumn} to defer loading sensitive fields, we mistakenly listed the column using its query method name \texttt{exam?} instead of its actual name \texttt{exam}.
This caused the gem to override \texttt{exam?} to always return \texttt{nil} (which is falsy), silently disabling the exam-check logic.
So a correct-looking access check was rendered a no-op due to a misuse of an external library elsewhere in the codebase---%
a subtle bug that we discovered only after reviewing the policy extracted by \name.

\subsection{Broadening the Extracted Policy}\label{sec:eval:broaden}
Sometimes, an extracted policy includes many combinations of conditions under which data can be accessed,
reflecting business logic rather than privacy concerns.
For example, the extracted \diaspora policy has 39~views
(out of \diasporaFinalViews)
that reveal a profile under various conditions:
if it belongs to the current user, or to the author of a public post,
etc.
But based on our understanding,
a \diaspora user's profile is intended to be public, except for a few columns guarded by the \texttt{public\_details} flag.
This means we can simplify the policy by \emph{broadening} it.

We can broaden the policy with the help of \name's view pruning (\Cref{sec:policy:prune}).
First, we write down four simpler, broader views to capture the relaxed conditions for accessing profiles:
\begin{enumerate}[itemsep=3pt]
  \item
    \begin{compactsql}
SELECT id, first_name, ... FROM profiles
    \end{compactsql}
    \textit{(For all profiles, some columns are always visible\ldots)}
  \item
    \begin{compactsql}
SELECT * FROM tags, taggings
WHERE tags.id = taggings.tag_id
  AND taggings.taggable_type = 'Profile';
    \end{compactsql}
    \textit{(\ldots and so are the profile taggings.)}
  \item
    \begin{compactsql}
SELECT * FROM profiles
WHERE public_details = TRUE
    \end{compactsql}
    \textit{(All columns are visible if the profile's ``public details'' flag is set\ldots)}
  \item
    \begin{compactsql}
SELECT * FROM profiles, people
WHERE profiles.person_id = people.id
  AND people.owner_id = MyUserId
    \end{compactsql}
    \textit{(\ldots or if the profile belongs to the current user.)}
\end{enumerate}
We add these to the policy and re-run view-pruning,
which removes 36~of the 39 profile-related views as redundant.
This process spares the human from having to reason about view subsumption, which can be tedious and tricky.
(For example, the remaining three views, which pertain to the current user's contacts, are \emph{not} subsumed by the ones that we added!)

\subsection{Manual Effort and Expertise Required}\label{sec:eval:manual-effort}
The steps involved in applying \name required varying degrees of SQL, Rails, and application knowledge.
Before extraction:
\begin{itemize}
  \item The \emph{code changes} (\Cref{sec:eval:app-setup:code}) were mostly compatibility edits that required general Rails and SQL familiarity.
    The only application-specific changes---fetching sensitive data only when it affects output---were inherited from our prior work and are not required by \name.
  \item Writing \emph{relevance hints} (\Cref{sec:eval:app-setup:hints}) required Rails knowledge to determine whether particular expressions or library calls could issue queries.
  \item Writing manual \emph{database constraints} (\Cref{sec:eval:app-setup:constraints}) required familiarity with the application's data model, including its schema, Rails model associations and validators, and invariants encoded only in application logic.
\end{itemize}
After extraction:
\begin{itemize}
  \item \emph{Query approximations} (\Cref{sec:eval:stats}) required SQL knowledge to assess their effect on the policy, and application familiarity to decide whether they were appropriate.
  \item \emph{Policy review} (\Cref{sec:eval:compare}) required application and SQL knowledge to explain policy differences as policy mistakes, application bugs, or benign discrepancies.
  \item \emph{Policy broadening} (\Cref{sec:eval:broaden}) required application familiarity to identify non-privacy-relevant distinctions, and SQL knowledge to write the broader views.
\end{itemize}

%% file: figures/stats-macros.tex
\newcommand{\diasporaFinalViews}{\num{134}\xspace}
\newcommand{\autolabFinalViews}{\num{144}\xspace}
\newcommand{\theodinprojectFinalViews}{\num{24}\xspace}

%% file: figures/main-table.tex
\begin{NiceTabular}{
        l 
        r 
        r 
        @{\hskip 0.5\tabcolsep}c@{\hskip 0.5\tabcolsep}
        @{\hskip 0pt}r 
        r 
        l 
        r 
        r 
        r 
        r 
        r 
        }
        \toprule
        & \Block{1-6}{\textbf{Statistics}} & & & & & & \Block{1-5}{\textbf{Running Time}} & & & & \\
        \cmidrule(lr){2-7} \cmidrule(lr){8-12}
        \textbf{Handler} & \#Paths & \Block{1-3}{\#Cond.~Queries} & & & \Block{1-2}{\#SQL Views} & & Explore & Simplify & Prune & Final Prune & \textbf{Total} \\
\midrule
\textbf{\diaspora} & & & & & & & & & & \SI{36}{\minute} & \textbf{\SI{4.7}{\hour}} \\
\quad\texttt{People\#stream} \prune{} & \num{163832} & \num{189412} & $\to$ & \num{188} & \num{148} &\Block{6-1}{$\to$ \num{134}}& \SI{1.1}{\hour} & \SI{8}{\minute} & \SI{35}{\minute} & & \\
\quad\texttt{Posts\#show} \prune{} & \num{73489} & \num{47256} & $\to$ & \num{208} & \num{87} && \SI{32}{\minute} & \SI{1}{\minute} & \SI{20}{\minute} & & \\
\quad\texttt{People\#show} \prune{} & \num{27480} & \num{1318} & $\to$ & \num{35} & \num{27} && \SI{15}{\minute} & \SI{33}{\second} & \SI{4}{\minute} & & \\
\quad\texttt{Notifications\#index} \prune{} & \num{32907} & \num{1018} & $\to$ & \num{75} & \num{54} && \SI{16}{\minute} & \SI{38}{\second} & \SI{9}{\minute} & & \\
\quad\texttt{Conversations\#index} & \num{57534} & \num{93368} & $\to$ & \num{134} & \num{39} && \SI{14}{\minute} & \SI{4}{\minute} & \SI{12}{\minute} & & \\
\quad\texttt{Comments\#index} & \num{3607} & \num{41} & $\to$ & \num{23} & \num{17} && \SI{2}{\minute} & \SI{24}{\second} & \SI{3}{\minute} & & \\
\midrule
\textbf{Autolab} & & & & & & & & & & \SI{29}{\minute} & \textbf{\SI{3.9}{\hour}} \\
\quad\texttt{Assessments\#show} \prune{} & \num{217543} & \num{391621} & $\to$ & \num{1592} & \num{138} &\Block{6-1}{$\to$ \num{144}}& \SI{1.9}{\hour} & \SI{28}{\minute} & \SI{32}{\minute} & & \\
\quad\texttt{Assessments\#gradesheet} & \num{15681} & \num{9020} & $\to$ & \num{46} & \num{35} && \SI{6}{\minute} & \SI{1}{\minute} & \SI{6}{\minute} & & \\
\quad\texttt{Assessments\#index} & \num{7839} & \num{9993} & $\to$ & \num{86} & \num{27} && \SI{4}{\minute} & \SI{50}{\second} & \SI{7}{\minute} & & \\
\quad\texttt{Submissions\#download} & \num{79} & \num{53} & $\to$ & \num{21} & \num{18} && \SI{1}{\minute} & \SI{21}{\second} & \SI{3}{\minute} & & \\
\quad\texttt{Courses\#index} & \num{36} & \num{17} & $\to$ & \num{8} & \num{5} && \SI{2}{\minute} & \SI{21}{\second} & \SI{50}{\second} & & \\
\quad\texttt{Metrics\#getNumPending} & \num{11} & \num{14} & $\to$ & \num{8} & \num{7} && \SI{57}{\second} & \SI{20}{\second} & \SI{59}{\second} & & \\
\midrule
\textbf{The Odin Project} & & & & & & & & & & \SI{8}{\minute} & \textbf{\SI{59}{\minute}} \\
\quad\texttt{Lessons\#show} & \num{48670} & \num{49459} & $\to$ & \num{152} & \num{46} &\Block{6-1}{$\to$ \num{24}}& \SI{13}{\minute} & \SI{3}{\minute} & \SI{13}{\minute} & & \\
\quad\texttt{Courses\#show} & \num{4462} & \num{3133} & $\to$ & \num{37} & \num{25} && \SI{3}{\minute} & \SI{30}{\second} & \SI{4}{\minute} & & \\
\quad\texttt{ProjectSubmissions\#index} & \num{3052} & \num{2615} & $\to$ & \num{39} & \num{17} && \SI{3}{\minute} & \SI{31}{\second} & \SI{3}{\minute} & & \\
\quad\texttt{Users\#show} & \num{2003} & \num{2059} & $\to$ & \num{14} & \num{9} && \SI{2}{\minute} & \SI{29}{\second} & \SI{1}{\minute} & & \\
\quad\texttt{Paths\#index} & \num{28} & \num{26} & $\to$ & \num{6} & \num{4} && \SI{2}{\minute} & \SI{19}{\second} & \SI{38}{\second} & & \\
\quad\texttt{Sitemap\#index} & \num{28} & \num{13} & $\to$ & \num{4} & \num{3} && \SI{1}{\minute} & \SI{20}{\second} & \SI{27}{\second} & & \\
\bottomrule
\CodeAfter\SubMatrix.{4-6}{9-6}\}
\CodeAfter\SubMatrix.{11-6}{16-6}\}
\CodeAfter\SubMatrix.{18-6}{23-6}\}
\end{NiceTabular}

%% file: figures/relevance-table.tex
\begin{tabular}{
        l 
        r 
        r 
        r 
        }
        \toprule
        & \multicolumn{2}{c}{\textbf{Counts}} & \textbf{Duration (min)} \\
        \cmidrule(lr){2-3}
        \textbf{Handler} & Rel. & Irrel. & (mean $\pm$ std) \\
\midrule
\textbf{\diaspora} & & & \\
\quad\texttt{People\#stream} & \num{31} & \num{242} & $\num{1.3} \pm \num{0.4}$ \\
\quad\texttt{Posts\#show} & \num{13} & \num{171} & $\num{1.4} \pm \num{0.4}$ \\
\quad\texttt{People\#show} & \num{16} & \num{86} & $\num{1.6} \pm \num{0.4}$ \\
\quad\texttt{Notifications\#index} & \num{21} & \num{82} & $\num{1.9} \pm \num{0.7}$ \\
\midrule
\textbf{Autolab} & & & \\
\quad\texttt{Assessments\#show} & \num{46} & \num{54} & $\num{1.0} \pm \num{0.3}$ \\
\bottomrule
\end{tabular}

%% file: related.tex
\paragraph{Symbolic execution}
Symbolic execution~\cite{DBLP:journals/cacm/King76} is a classic path-exploration technique that maintains symbolic program state.
We decided not to use it because implementing it for an interpreted language is challenging (more so than for lower-level languages~\cite{DBLP:conf/osdi/CadarDE08,DBLP:conf/kbse/MossbergMHGGFBD19,DBLP:conf/sp/Shoshitaishvili16})
due to the dynamic features and functionality implemented outside the language~\cite[\S~2.2]{DBLP:conf/asplos/BucurKC14}.

\paragraph{Bug finding}
Many tools use symbolic or concolic execution to find bugs in web applications~\cite{DBLP:conf/ccs/ChaudhuriF10,DBLP:conf/issta/WassermannYCDIS08,DBLP:conf/issta/ArtziKDTDPE08},
typically aiming to trigger certain statements or states.
Policy extraction requires not only reaching the query-issuing statements,
but also gathering the conditions under which they are reached.
Other systems infer security-relevant behavior from code to discover violations.
AutoISES~\cite{DBLP:conf/uss/TanZMXZ08} infers expected security checks around OS operations and reports omissions;
iHunter~\cite{DBLP:conf/uss/LiuXZXBZX24} finds privacy violations in iOS SDKs by recovering privacy-sensitive data flows;
and Derailer~\cite{DBLP:conf/kbse/NearJ14} and Space~\cite{DBLP:conf/icse/NearJ16} check for security bugs in web applications by validating data exposures against human input or a catalog.
Such tools use inferred behavior as evidence of bugs, whereas \name produces an access-control policy for review and enforcement.

\paragraph{Syscall filtering}
Abhaya~\cite{DBLP:journals/pacmpl/Pailoor0SD20} synthesizes Seccomp-bpf and Pledge policies by statically analyzing a program's syscall behavior and finding a tight policy expressible in the target sandbox language.
SysPart~\cite{DBLP:conf/ccs/RajagopalanKGXP23}, C2C~\cite{DBLP:conf/ccs/GhavamniaPP22}, and Sysfilter~\cite{DBLP:conf/raid/DeMarinisWJFK20} similarly harden binaries by inferring the set of syscalls a program may invoke and restricting the allowed syscalls to this set.
These systems are conceptually similar to \name in that they infer policies from programs.
However, they typically rely on static analysis (sometimes complemented by dynamic observations) and reason about a smaller policy space: syscall identities and, in some systems, predicates over syscall arguments.
In contrast, \name targets a dynamic scripting language and extracts expressive relational database policies, including the conditions under which SQL queries are issued.

\paragraph{Instrumentation strategies}
Some systems track symbolic values in
dynamic languages by performing instrumentation \emph{within} the target language.
They represent symbolic objects through proxying~\cite{BruniDF11,DBLP:conf/uss/WangNXC13} or inheritance~\cite{DBLP:series/natosec/BallD15}, and track path conditions using Boolean-conversion hooks~\cite{BruniDF11,DBLP:series/natosec/BallD15} or debug tracing~\cite[\S~4.1.2]{DBLP:conf/uss/WangNXC13}.
These approaches avoid the need for a custom interpreter and support standard environments~\cite{DBLP:conf/cloud/AlpernasPRS21}.
In contrast, \name performs \emph{offline} analysis, allowing us to modify the interpreter for better transparency~\cite{crosshair-limitations} and performance.

\paragraph{Learning models of applications}
Konure infers models of database-backed applications~\cite{DBLP:journals/toplas/ShenR21,DBLP:conf/pldi/0001R19} by
generating targeted inputs to probe the application as a black-box.
This work informed our formulation of the simple query-issuing core assumption (\Cref{sec:concolic:simple}).
However, black-box probing cannot effectively recover conditions of query issuance~\cite[\S~6.2]{DBLP:journals/toplas/ShenR21}, whereas \name can extract them directly via instrumentation.

\paragraph{Policy mining}
Policy-mining systems share our goal of generating access-control rules that capture existing practices.
Most take existing access-control lists~\cite{DBLP:journals/corr/XuS13,DBLP:conf/sacmat/XuS12,DBLP:conf/sacmat/BuiS20}, operation logs~\cite{DBLP:conf/trustbus/Gal-OzGYGRS11,DBLP:conf/sacmat/MolloyPC12,DBLP:conf/sacmat/IyerM18}, or human interactions~\cite{DBLP:conf/sacmat/Iyer023},
and produce role-~\cite{DBLP:conf/sacmat/XuS12,DBLP:conf/trustbus/Gal-OzGYGRS11,DBLP:conf/sacmat/MolloyPC12}, attribute-~\cite{DBLP:journals/corr/XuS13,DBLP:conf/dbsec/XuS14,DBLP:conf/sacmat/IyerM18}, or relationship-based~\cite{DBLP:conf/sacmat/Iyer023,DBLP:conf/sacmat/BuiS20} rules, often via statistical techniques.
Unlike these systems, \name does not require live data and instead analyzes the application, which provides the visibility needed to produce fine-grained policies.
AutoArmor~\cite{DBLP:conf/uss/LiCLWC21} is closer to \name in that it analyzes application code to generate access-control policies, but it targets inter-service calls in microservice applications whereas \name extracts database access-control policies from web applications.

%% file: conclusion.tex
\name enables better data security in existing web applications:
by extracting the access-control policy embedded in application code, it makes that policy explicit, allowing an admin to understand and ultimately enforce the intended policy.

Although \name is semi-automated and provides no formal guarantees, these traits are shared by most policy-assistance tools.
RBAC and ABAC policy-mining tools (\Cref{sec:related}) routinely depend on human-curated inputs~\cite{DBLP:journals/tissec/FrankBB13}, require humans in the loop~\cite[\S~7.2]{DBLP:journals/csur/MitraSVA16}, and offer no guarantees as they operate on partial~\cite{DBLP:journals/tdsc/KarimiAJA22,DBLP:conf/eurosp/CotriniWB18,DBLP:conf/sacmat/MolloyPC12} or noisy~\cite{DBLP:conf/sacmat/NaroueiT15,DBLP:conf/sacmat/AlohalyTB18,DBLP:conf/bibm/XiaZWHWS22} data using statistical methods.
Yet these tools have consistently proven effective in helping admins develop practical policies~\cite{DBLP:journals/csur/MitraSVA16,DBLP:journals/csur/JayasundaraAR25,DBLP:journals/corr/abs-2207-01739,Tsarynny22,tufin15}.
These broader successes, together with our own, give us confidence that \name can provide similar---if not greater---value, particularly because \name generates far finer-grained policies than typical RBAC or ABAC rules.

We conclude with a few avenues for future work:

\paragraph{Automation bias and complacency}
A policy extractor may create a false sense of security when reviewers accept a plausible-looking policy without sufficient scrutiny.
This risk is an instance of \emph{automation bias and complacency}: Reviewers may over-rely on computer output and abdicate their decision-making responsibility~\cite{automation}.
Future work should investigate these effects in the context of policy extraction and adapt established mitigations to this setting~\cite{goddard2012automation}.

\paragraph{Policy comprehension}
SQL is precise and familiar, but SQL views can become verbose when they encode many application-level conditions.
A natural next step is a concise policy domain-specific language~(DSL) that preserves SQL-view semantics while making policies easier to read, compare, and audit.
Such a DSL should draw on query-language usability studies~\cite{DBLP:conf/iticse/AhadiPBL15,DBLP:journals/csur/Reisner81} and ORM abstractions~\cite{DBLP:journals/infsof/TorresGPM17};
verified lifting techniques~\cite{DBLP:conf/pldi/KamilCIS16,DBLP:journals/pacmpl/LaddadPMCH22} could translate generated SQL views into this DSL while preserving their meaning.

\paragraph{Coding agents}
The rapid improvement of coding agents raises a natural question: Could an agent simply read source code and extract a policy directly?
In principle, yes---human experts can do this, albeit slowly and imperfectly, and a capable enough agent might automate it.
But the goal is not merely to produce a plausible policy; it is to produce one whose provenance a human can audit and understand.

\name has limitations: it applies only under stated assumptions and relies on certain heuristics.
But these limitations are explicit.
\name decomposes policy extraction into well-defined subproblems grounded in concrete traces, derives policies through explicit transformations, and produces intermediate artifacts that help reviewers trace a policy fragment to its source.%
\footnote{Even when \name uses an LLM-based relevance judge, it records the judge's verdicts and explanations for audit.}
An end-to-end agent may also make mistakes, but those mistakes can be harder to localize because they arise from an opaque reasoning process with unclear failure modes.

Rather than replacing \name's structure, agents could operate within it: modeling complex application logic, flagging suspicious views, suggesting policy broadenings, etc.
This way, agents are used for well-scoped subtasks whose outputs are attached to explicit artifacts, so that increased automation does not come at the cost of auditability.

%% file: artifact-appendix.tex
\twocolumn[
\begin{center}
\captionof{table}{Artifact hosting locations.}\label{tbl:artifact-hosting}
\small\ra{1.1}
\begin{tabular}{lll}
\toprule
\textbf{Content} & \textbf{Location} & \textbf{Branch / release} \\
\midrule
\textbf{Artifact README} &
\url{https://github.com/ote-project/artifact-eval} &
\texttt{main} branch \\

\textbf{\name implementation} \\
\quad Core implementation (\Cref{sec:concolic,sec:policy}) &
\url{https://github.com/ote-project/concolic_driver} &
\texttt{main} branch \\
\quad Modified JRuby (\Cref{sec:concolic:instrumentation}) &
\url{https://github.com/ote-project/jruby} &
\texttt{9.3.13.0-ote} branch \\
\quad Modified Blockaid (\Cref{sec:policy:prune}) &
\url{https://github.com/ote-project/blockaid} &
\texttt{standalone-checking} branch \\

\textbf{Applications} (\Cref{sec:eval:app-setup}) \\
\quad \diaspora & \url{https://github.com/ote-project/diaspora} & \texttt{ote-v0.7.14.0} branch \\
\quad Autolab & \url{https://github.com/ote-project/Autolab} & \texttt{ote-v2.7.0} branch \\
\quad The Odin Project & \url{https://github.com/ote-project/theodinproject} & \texttt{ote-ruby2.6} branch \\

\textbf{Experiments} \\
\quad Virtual-machine image & \url{https://github.com/ote-project/artifact-eval} & \href{https://github.com/ote-project/artifact-eval/releases/tag/ae}{\texttt{ae} release tag (assets)} \\
\quad Experiment scripts & \url{https://github.com/ote-project/ae-scripts} & \texttt{main} branch \\
\quad Application config for \name (\Cref{sec:overview:workflow}) & \url{https://github.com/ote-project/ae-app-config} &
\texttt{main} branch \\
\quad Handwritten policy baseline (\Cref{sec:eval:compare}) & \url{https://github.com/ote-project/app-policies} & \texttt{ote} branch \\
\bottomrule
\end{tabular}
\end{center}
\vspace{1em}
]

\section{Artifact Appendix}
\subsection*{Abstract}

This artifact provides the \name implementation, including the concolic-execution driver and supporting modifications to JRuby and Blockaid, along with the modified applications and setup needed to reproduce the evaluation.

\subsection*{Scope}

The artifact reproduces \Cref{tbl:main,tbl:relevance,tbl:compare} in a PDF and writes the extracted policies to separate output files.

\subsection*{Contents}

The artifact includes the source repositories for \name's concolic-execution driver (\Cref{sec:concolic:architecture}), the JRuby interpreter modified for symbolic tracking (\Cref{sec:concolic:instrumentation}), Blockaid modified to support view pruning (\Cref{sec:policy:prune}), and the three modified applications used in the evaluation (\Cref{sec:eval:app-setup}).
It also provides a prebuilt virtual-machine image with the scripts and configuration files needed to reproduce the evaluation.

\name uses an LLM-based relevance judge implemented with OpenAI's Codex CLI (\Cref{sec:eval:experiment-setup}).
To save resources, the artifact includes a ``mock judge'' that, instead of calling an LLM, returns a verdict after a configured delay.
By default, the artifact's experiment script uses the mock judge with a 90-second per-call delay, approximating the real judge's latency.

\subsection*{Hosting}

The artifact is hosted under the \texttt{ote-project} GitHub organization
(\url{https://github.com/ote-project}). \Cref{tbl:artifact-hosting} lists the
hosting location for each component.

\subsection*{Requirements}

The artifact was tested on Google Cloud using a Compute Engine \texttt{c3-standard-176} instance.
Importing the VM image requires access to a Google Cloud Storage bucket for uploading the image before import.
Running \name with the real Codex-based relevance judge requires an OpenAI account.